\documentclass[a4paper]{jpconf}
\usepackage{graphicx}
\begin{document}
\title{Quasi-normal frequencies: Semi-analytic results for highly damped modes}

\author{Jozef Skakala and Matt Visser}

\address{School of Mathematics, Statistics and Operations Research,\\
 Victoria University of Wellington,\\
 New Zealand}

\ead{jozef.skakala@msor.vuw.ac.nz, matt.visser@msor.vuw.ac.nz}

\begin{abstract}
Black hole highly-damped quasi-normal frequencies (QNFs) are very often of the form  $\omega_n =  \hbox{(offset)} + i n \; \hbox{(gap)}$. We have investigated the genericity of this phenomenon for the Schwarzschild--deSitter (SdS) black hole by considering a model potential that is piecewise Eckart (piecewise P\"oschl--Teller), and developing an analytic ``quantization condition'' for the highly-damped quasi-normal frequencies. We find that the $\omega_n =  \hbox{(offset)} + i n \; \hbox{(gap)}$ behaviour is \emph{common but not universal}, with the controlling feature being whether or not the ratio of the surface gravities is a rational number. We furthermore observed that the relation between rational ratios of surface gravities and periodicity of QNFs is very generic, and also occurs within different analytic approaches applied to various types of black hole spacetimes. These observations are of direct relevance  to any physical situation where highly-damped quasi-normal modes are important. 
\end{abstract}

\def\sech{{\mathrm{sech}}}
\def\ln{{\mathrm{ln}}}
\def\d{{\mathrm{d}}}

\section{Introduction}
Black hole quasi-normal modes (QNMs) become (after some suitable damping timescale) the dominant contribution arising from arbitrary perturbations of any black hole spacetime. Thus they form the black hole's characteristic ``sound''.
They are solutions of Regge--Wheeler (axial), or Zerelli (polar) equation, with outgoing radiation boundary conditions. The Regge--Wheeler (Zerelli) equation becomes a time-independent Schrodinger-like equation for the ``eigenvalues'' $\omega_n^{2}$, where the $\omega_n$ are non-real quasi-normal frequencies. The real parts of the QNM frequencies are spaced symmetrically with respect to the imaginary axis, and with suitable conventions the imaginary part is always positive. The background black hole spacetime is encoded in the Regge--Wheeler (Zerelli) potential, which typically shows exponential fall-off as one moves towards the event horizon(s). The rate of this fall-off is given by the horizon's surface gravity. 

Originally, interest was focused on the fundamental lowest damped modes. One of the methods to calculate them was to approximate the real potential by the analytically solvable P\"oschl--Teller potential \cite{Ferrari}. To obtain the fundamental modes one needs to fit the P\"oschl--Teller potential by the real potential's peak height and peak curvature. 

Interest in the highly damped modes ($Im(\omega)$ being large) significantly increased after Hod's conjecture about their relation to black hole thermodynamics was formulated \cite{Hod}. There are good physics arguments to expect that the highly damped modes depend qualitatively on the behavior of the tails of the potential. (The wave-pocket formed from such modes spreads quickly to the region of the tails of the potential.) Thus approximating the tails of the potential by analytically solvable potentials is expected to bring considerable \emph{qualitative} understanding regarding the behavior of highly damped QNMs. Perhaps surprisingly, such an approximation is both tractable and informative, at least for the Schwarzschild spacetime and Schwarzschild--de Sitter spacetime. Detailed calculations are contained in \cite{Skakala1, Skakala2} and the basic results are summarized in this paper.

\section{Schwarzschild spacetime}

There is a wide agreement regarding the behavior of asymptotic (highly damped) quasi-normal modes in the case of Schwarzschild asymptotically flat spacetime \cite{review, Nollert}. They are described by the formula (see for instance \cite{Motl2, Medved1, Medved2}): 
\begin{equation}
\omega=\frac{\ln 3}{2\pi}\kappa +i\kappa \left(n+\frac{1}{2}\right)+O(n^{-1/2}).\label{asymptotic}
\end{equation}
In this case the tails of the Regge--Wheeler (Zerelli) potential are approximated by the exactly solvable piecewise smooth potential: 
\begin{equation}
V(x) = \left\{  \begin{array}{lcl}
V_{0-} \; \sech^2(\kappa x)  & \hbox{ for }  & x < 0. \\
\\
{V_{0+}/(x+a)^{2}}  & \hbox{ for }  & x > 0, \quad a > 0;\\
\end{array}\right.
\end{equation}
where $\kappa$ is the surface gravity of the black hole horizon. For $x>0$ the solutions giving the outgoing radiation (for $V_{0+}\geq-1/4$) are~\cite{Skakala1, Skakala2}
\begin{equation}
\psi_{+}(x)=C_{+}\; \sqrt{\frac{x+a}{\omega}}\; \left[J_{\alpha_{+}}(\omega(x+a))-e^{-i\alpha\pi}J_{-\alpha_{+}}(\omega(x+a)\right], \qquad(x > 0),
\end{equation}
where the $J_{\alpha_{+}}(x)$ are Bessel functions with
\begin{equation}
\alpha_{+}\doteq\sqrt{1+4V_{0+}}.
\end{equation}
On the other hand, for $x<0$ we have~\cite{Skakala1, Skakala2}
\begin{equation}
\psi_{-}(x) = C_{-}\; e^{i\omega x} \; {} _2F_1\left({1\over2}+\alpha_{-},{1\over2}-\alpha_{-},1+i\frac{ \omega}{\kappa}, {1\over 1+ e^{- 2\kappa x}} \right),
\qquad (x < 0),
\end{equation}
where ${}_2F_1(...)$ is the hypergeometric function with
\begin{equation}
\alpha_- =
\sqrt{{1\over4} - {V_{0-}\over\kappa^2}}
\end{equation}
The equation for QNM modes is obtained from the junction condition:
\begin{equation}
\frac{\psi'_{+}(0)}{\psi_{+}(0)}=\frac{\psi'_{-}(0)}{\psi_{-}(0)}.\label{flat}
\end{equation}
The derivatives of both hypergeometric and Bessel functions can be rewritten in terms of hypergeometric and Bessel functions themselves. Furthermore, at the specific point of interest, the hypergeometric functions can themselves be rewritten in terms of Gamma functions, but it is still difficult to look for solutions of (\ref{flat}) without taking some approximations \cite{Skakala1, Skakala2}. Fortunately from the beginning of the problem we were interested only in highly damped modes, where $Im(\omega) \gg 0$ and $Im(\omega)\gg|Re(\omega)|$. In this case we can approximate both Bessel functions and also Gamma functions (after playing with some Gamma identites we can apply the Stirling formula) by trigonometric functions. Then (\ref{flat}) simplifies to the extremely simple condition:
\begin{equation}
\sin\left(\frac{\pi\omega}{\kappa}\right)=0.
\end{equation}
The solution of this is: $\omega=i\kappa n$, qualitatively recovering the result (\ref{asymptotic}) at leading order.

\section{Schwarzschild--deSitter spacetime}
In this situation the exactly solvable potential approximating the tails of Regge--Wheeler (Zerelli) potential is~\cite{Skakala1, Skakala2} (see also~\cite{Suneeta}):
\begin{equation}
V(x) = \left\{  \begin{array}{lcl}
{V_{0-} \; \sech^2(\kappa_- x)}  & \hbox{ for }  & x < 0;\\
\\
V_{0+} \; \sech^2(\kappa_+ x)  & \hbox{ for }  & x > 0. \\
\end{array}\right.
\end{equation}
Here $\kappa_\pm$ again denotes the surface gravities of the two horizons (black hole and cosmological). The solutions of the Regge--Wheeler (Zerelli) equation are
\begin{equation}
\psi_{\pm}(x) = C_{\pm}\; e^{\mp i\omega x} \; {} _2F_1\left({1\over2}+\alpha_{\pm},{1\over2}-\alpha_{\pm},1+i\frac{ \omega}{\kappa_{\pm}}, {1\over 1+ e^{- 2\kappa_{\pm} x}} \right),
\end{equation}
where
\begin{equation}
\alpha_\pm =
\sqrt{{1\over4} - {V_{0\pm}\over\kappa^2_{\pm}}}
\end{equation}
Now after applying the junction condition (\ref{flat})
we can proceed in the same way as previously. (Rewriting derivatives of hypergeometric functions in terms of hypergeometric functions, evaluating the hypergeometric functions at the particular point of interest in terms of Gamma functions, using Gamma function identities and the Stirling approximation 
for $Im(\omega) \gg 0$ and $Im(\omega) \gg |Re(\omega)|$.) After some calculations \cite{Skakala1, Skakala2} we obtain:
\begin{equation}
 \label{qnf}
\cos(-i\pi\omega[\kappa^{-1}_+ - \kappa^{-1}_-]) - \cos(-i\pi\omega[\kappa^{-1}_+ + \kappa^{-1}_-]) = 2  \;  \cos(\pi\alpha_+)\cos(\pi\alpha_-).
 \end{equation}
The solutions of equation (\ref{qnf}) have numerous interesting properties. In particular:
\newtheorem{one}{Theorem}[section]
\begin{one}
Take the case when $\kappa_+/\kappa_-=p_-/p_+\in\mathbf{Q}$. Take $\kappa_*=p_\pm \kappa_\pm$ ($p_\pm\in\mathbf{N}$), hence $\kappa_*=lcm(\kappa_+ ,\kappa_-)$. Then the solutions of (\ref{qnf}) are given by equi-spaced families of quasi-normal modes: 
\begin{equation}
\omega_{an}=\omega_{0 a}+ing\kappa_*~~~~ (n\in\mathbf{N}).
\end{equation} 
Here $g=1$ if $p_+\cdot p_-$ is odd and $g=2$ if $p_+\cdot p_-$ is even. The $\omega_{a}= \frac{\kappa_* \ln(z_a)}{\pi}$ are related to roots $z_a$ of the  specific polynomial:
\begin{equation}
z^{2[p_+ +p_-]} - z^{2p_+} -  z^{2p_-} -
4\cos\left(\pi\alpha_+\right) \cos\left(\pi\alpha_-\right) z^{[p_+ +p_-]}  + 1.\label{polynomial}
\end{equation}
\label{One}
\end{one}
\newtheorem{two}[one]{Theorem}
\begin{two}
Assume that (\ref{qnf}) has some set of solutions of the form $\omega_{n}=\omega_{0}+in\cdot (gap)$ ($n\in\mathbf{N}$). Then the ratio $\kappa_+/\kappa_-$ must be rational.\label{Two}
\end{two}
\enlargethispage{20pt}
Thus existence of infinite equi-spaced sets of solutions of the equation (\ref{qnf}) is \emph{equivalent} to the rational ratio of surface gravities. Whenever the ratio is rational, the generic number of different equi-spaced families is $g(p_+ + p_-)$, as it is related to the number of roots of the polynomial (\ref{polynomial}). 

\section{Discussion}
We have presented strong physics arguments why our model is likely to give good \emph{qualitative} estimates for the QNFs of the real Regge-Wheeler potential of the SdS spacetime. But ``\emph{qualitative}'' means that only the periodicity with the correct gap structure is recovered, the modes do not have to necessarily \emph{quantitatively} match. (Specifically, the $\omega_{0}$ mode might differ.) Now, since different families of modes are distinguished only by different $\omega_{0a}$ the following question can be asked: Does the multi-family splitting transfer from our approximate model to the real SdS case? The answer to this question is suggested by the following statement (which can be exactly proven): 

\emph{Assume that the gap structure is of the form given in (\ref{One}). Assume also that the number of families is a bounded function of the variables $\kappa_+, \kappa_-$ ($\kappa_+/\kappa_-\in\mathbf{Q}$). Then for every point $\tilde\kappa_+, \tilde\kappa_-$ it holds that maximally one QNF can be a continuous function of $\kappa_+, \kappa_-$ variables at the point $\tilde\kappa_+, \tilde\kappa_-$.}

This statement strongly suggests (since there are good reasons to expect that QNMs are continuous functions of surface gravities) that if the gap structure transfers to the real physical situation, also the basic features of multi-family splitting must transfer. 
Furthermore \cite{Skakala1} we are able to show that the limit $\kappa_+\to 0$ gives in our model the same results as the case $V_{0+}=0$. In both cases we obtain the QNFs of Schwarzschild spacetime, which is the expected result.

A complementary set of semi-analytic results for QNFs is obtained by using monodromy techniques \cite{review, Motl2, NS, Cardoso, Das}. All the equations derived by this method (for Schwarzschild, Schwarzschild--de Sitter, Reissner--Nordstr\"om, Reissner--Nordstr\"om--de Sitter) are of the form~\cite{Skakala3}
\begin{equation}
\sum_{n=1}^{N}{A_{n}\exp\left(i\pi\omega\sum_{j=1}^{H}\frac{B_{nj}}{\kappa_{j}}\right)}=0.
\end{equation}
As a result of this, the theorem \ref{One} can be completely generalized for all the known analytic results, and the theorem \ref{Two} is very generic as well (it holds with very rare and mostly unphysical exceptions) \cite{Skakala3}.

\end{document}